\documentclass{aa}
\usepackage{epsfig}
\def\ltsima{$\; \buildrel < \over \sim \;$}
\def\simlt{\lower.5ex\hbox{\ltsima}}
\def\gtsima{$\; \buildrel > \over \sim \;$}
\def\simgt{\lower.5ex\hbox{\gtsima}}
\def\src{X{\thinspace}1724$-$308}

\begin{document}
\thesaurus{}
   \thesaurus{6(13.25.5;  % X-rays: stars,
		10.07.3 Terzan~2; % globular cluster: individual:
                08.09.2;  % stars: individual: \src,
                08.14.1;  % stars: neutron,
                08.02.1;  % binaries: close,
                02.01.2)}  % accretion: accretion discs

\title{The Comptonized X-ray source \src\ in the globular cluster Terzan~2}

\author{M. Guainazzi\inst{1}, A.N. Parmar\inst{1}, A. Segreto\inst{1,2}, L. Stella\inst{3}, D. Dal Fiume\inst{4},
T.Oosterbroek\inst{1}}

\institute{
{Astrophysics Division, Space Science Department of ESA, ESTEC, Postbus 299,
NL-2200 AG Noordwijk, The Netherlands}
\and
{Istituto di Fisica Cosmica ed Applicazioni dell'Informatica, C.N.R., Via Ugo la Malfa 153, I-90146 Palermo, Italy}
\and
{Osservatorio Astronomico di Roma, Via dell'Osservatorio, I-00044 Monteporzio Catone, Italy}
\and
{Istituto Tecnologie e Studio delle Radiazioni Extraterrestri, C.N.R., Via Gobetti 101, I-40129 Bologna, Italy}
}
   
\offprints{M.Guainazzi [mguainaz@astro.estec.esa.nl]}

\date{Received 16 June 1998 ; accepted 31 August 1998}

\maketitle

\markboth{M.Guainazzi et al.}{Evidence of Comptonization from the broadband spectrum and variability of the X--ray source in Terzan~2}

\begin{abstract}

We report on the BeppoSAX observation of the X-ray source \src\
in the globular cluster
Terzan~2. The broadband spectrum can be described
as the superposition
of a power--law with photon spectral index $\Gamma \simeq 1.6-1.9$
and a thermal component
with typical temperature ${\rm kT \sim}$1~keV. \src\ is detected in
the PDS up to 150~keV with a S/N ratio $>$3,
notwithstanding a sharp exponential cutoff
with ${\rm E_{cutoff} \simeq 90}$~keV. The broadband spectrum can be
best interpreted as due to the
Comptonization of a Wien photon distribution with ${\rm kT_W} \sim$1~keV
by a spherically-symmetric hot plasma with electron temperature
${\rm kT_e \sim 30}$~keV and optical depth ${\rm \tau \simeq 3}$.
A comparison with the fainter and steeper state, which
was observed by EXOSAT, suggests a correlation between
the temperature of the Comptonizing electron distribution
(or the plasma optical depth) and the X-ray intensity.
The X--ray thermal component could originate from
a boundary layer between the accretion disk and the compact object,
or from the accretion disk itself.

\end{abstract}


\section{Introduction}

The globular cluster Terzan~2 has been known to contain a
luminous X-ray source since the late 1970s. An X-ray burst
from a region including the cluster was observed by OSO-8
(Swank et al. 1977), and the same mission detected persistent
emission from a location consistent with the cluster. 
{\it Uhuru} (Forman et al. 1978)
and {\it Einstein} High Resolution Imager (HRI, Grindlay et al. 1980) observations placed this source
in the core of the cluster. The source is known either
as \src\ (Parmar et al. 1989; hereafter P89) or 1E\thinspace1724$-$305.
It has 
a rather hard (photon index $\Gamma \sim$2) power--law spectrum, with a
1--20~keV flux which varies by about a factor of 4
($2.5-10 \times 10^{-10}$~erg~s$^{-1}$~cm$^{-2}$). {\it Einstein} data 
suggest that in a higher flux state
($F_{1-20 \ keV} \simeq 2.1 \times 10^{-9}$~erg~cm$^{-2}$~s$^{-1}$)
the source softens and the X-ray spectrum is better described
by a $\sim$6~keV bremsstrahlung (Barret et al. 1998, in preparation).

Terzan~2 is a relatively dim globular cluster, with an estimated distance
of $\sim$10~kpc and visual extinction ${\rm A_V = 4.03}$, according
to Djorgovski (1993). The scatter on the values reported in the literature
for these quantities is however rather high (${\rm \sim 30\%}$ and
${\rm \pm^{15}_5\%}$, respectively; see Ortolani et al. 1997 and
references therein).
The power--law intensity profile of its surface brightness and the relatively
low core star density argue for a complex dynamical history (Djorgovski
et al. 1986).

Terzan 2 is one of the two globular clusters from which X-ray
emission was detected at energies $>$40~keV by {\it Sigma}
(Barret et al. 1991).
The {\it Sigma} source (GRS\thinspace1724$-$308) was observed several times
between 1991 and 1994 and appears to be variable (Goldwurm et al. 1994).
The 1991 detections
(at a total 6.8$\sigma$ confidence level) are consistent with
a simple power--law spectrum with $\Gamma = 1.6^{+0.4}_{-0.6}$ and
a 38--200~keV flux of $\sim$$5.8 \times 10^{-3}$photons~cm$^{-2}$~s$^{-1}$,
corresponding to a poorly determined 1--20~keV flux of between
1.5 and $13 \times 10^{-10}$~erg~cm$^{-2}$~s$^{-1}$.
ROSAT HRI observations by Mereghetti et al. (1995)
did not reveal any other X-ray sources within 3\farcm5 from the
cluster core, supporting the identification of GRS\thinspace1724$-$308 
with X{\thinspace}1724$-$308.
The existence of a population a low-luminosity ({\it i.e.}: $\simlt$ a few
$10^{33}$~erg~s$^{-1}$) sources in Terzan~2,
similar to those discovered in the cores of
47~Tuc (Hasinger et al. 1994) or 
$\omega$~Cen (Johnston et al. 1994) cannot be excluded.
The X--ray flux of these sources would be $\sim$3
orders of magnitude lower than that of \src\ and
their existence therefore
would not affect any of the conclusions in this {\it paper}.

Recent RXTE observations of \src\ (Olive et al. 1998) suggest that
its timing properties are typical of an ``atoll''
source. The Power Spectrum Density (PSD) can be well modeled with
the sum of two ``shot noise'' components, with shot decay timescales
of $\sim$16 and $\sim$680~ms. A Quasi-Periodic Oscillation
(QPO) feature at $\sim$0.8~Hz was also detected in the PSD. The claim
of another QPO feature at ${\rm \nu \simeq 0.092}$~Hz observed by the
Medium Energy (ME) instrument on board EXOSAT (Belli et al. 1986)
remains unconfirmed. The X-ray spectral hardness and the continuum
properties of the PSD resemble those of some black hole candidates.

As part of the BeppoSAX Core Program, a number of X-ray luminous globular
clusters are being observed. A discussion of the
properties of the sample will be deferred to a forthcoming paper.
Here, we report the results of the BeppoSAX observation of \src\ . 

\section{Observation and data Reduction}

Results from all the co-aligned instruments on board BeppoSAX
(Boella et al. 1997a) are presented: the Low-Energy Concentrator Spectrometer (LECS;
0.1--10~keV; Parmar et al. 1997), the Medium-Energy Concentrator
Spectrometer (MECS; 1.3--10~keV; Boella et al. 1997b),
the High Pressure Gas Scintillation proportional Counter
(HPGSPC, 4--120~keV, Manzo et al. 1997)
and the Phoswich
Detection System (PDS; 15--300~keV; Frontera et al. 1997).
The MECS consisted at the time of the \src\ observation of three 
identical grazing incidence
telescopes with imaging gas scintillation proportional counters in
their focal planes (now only two are operating).
The LECS uses an identical concentrator system as
the MECS, but utilizes an ultra-thin (1.25~$\mu$m) entrance window and
a driftless configuration to extend the low-energy response to
0.1~keV.  The fields of view (FOV) of the LECS and MECS are circular
with diameters of 37\arcmin\ and 56\arcmin, respectively. The non-imaging
PDS consists of four independent units arranged in pairs each having a
separate collimator. Each collimator can be alternatively
rocked on- and off-source to monitor the background counting rate.
The hexagonal PDS FOV is 78\arcmin\ full-width at half maximum.
The HPGSPC was similarly operating in rocking mode, but only
the ``off-'' position has been employed to monitor the background.
Several pieces of evidence indicate that there are no significant
contaminating
sources present
in the PDS FOV, both in the ``on'' and in the ``off''
positions: the ``off'' count rates are in good agreement with the long--term
increasing trend of instrumental background studied by Guainazzi \& Matteuzzi
(1997); the ``on'' and ``off'' light curves show the same variability pattern
($\sim$20\% of count rates in the whole energy bandpass),
which is due to the variation of the particle background
around the orbit, thus excluding the presence of a strongly
variable or transient
unknown source and finally the PDS spectrum aligns very well with
those from the other instruments
(see Sect.~3). Similar arguments apply to the HPGSPC data.

The region of sky containing \src\ was observed by BeppoSAX
between 1996 August 17 04:29:05 UT and August 18 05:05:11 UTC in an
observation lasting 65~ks and spanning 12 orbital cycles.
Good data were selected from intervals when the elevation angle
above the Earth's limb was $>$$4^{\circ}$ and when the instrument
configurations were nominal, using the SAXDAS 1.3.0 data analysis package
(Lammers 1997; custom software has been used to reduce the HPGSPC data).
Additionally, PDS data within 5 minutes after each South Atlantic Anomaly
passage were excluded to avoid intervals were the gain was recovering to
its nominal value following switch-on.
This gives exposure times of 7.3~ks, 37.1~ks, 10.5~ks and 16.7~ks for the
LECS, MECS, HPGSPC, and PDS, respectively. 
The standard PDS (HPGSPC) collimator rocking angle of 210\arcmin\ (180\arcmin), and
standard dwell time of 96~s for each on- and off-source position were used.
The LECS was operated only during satellite night-time and was switched-off 
during the second half of the observation for technical reasons. 

\section{Data analysis}

LECS and MECS data were
extracted centered on the position of \src\ using radii of 8\arcmin\ and 
4\arcmin\, respectively, corresponding
to $\simeq 95\%$ of the instrumental Point Spread
Functions. Background subtraction in the imaging instruments
was performed using standard files, but is not critical for such a bright
source. Background subtraction in the HPGSPC and PDS was performed
using data from the 
offset detectors. The background-subtracted count rates of \src\ are
3.2~s$^{-1}$, 10.4~s$^{-1}$, 11.2~s$^{-1}$ and 6.1~s$^{-1}$ in the LECS, MECS, HPGSPC,
and PDS, respectively. Spectra have been rebinned so as to have at least
30 counts per energy channel and
to sample the instrumental energy resolution with
no more than 3 spectral channels, in order to ensure the applicability
of $\chi^2$ fitting techniques.
A systematic error of 1\% has been added to each
channel of the rebinned LECS and MECS spectra, to account for the residual
systematic uncertainties in the detector calibration.
Publicly available responses at 1997 September have been
used, and spectral fits performed in the following energy ranges:
0.1--4~keV (LECS), 2.2--10.5~keV (MECS), 4--40~keV (HPGSPC), 15--200~keV (PDS).
In the spectral fits, a normalization factor has been included to account for
the well-known mismatch in the BeppoSAX instrument
absolute flux calibration (Cusumano et al. 1998). The factors have been
left completely free, except for the PDS to MECS one, which has been
constrained to be in the 0.82-1.02 range (Cusumano et al. 1998).
The values of the LECS to MECS and HPGSPC to MECS relative normalizations
derived from the fit lie
in the intervals 0.80--0.82 and 1.05--1.06, respectively.
They are in good agreement with the values typically observed
(0.7--1.0 and 0.8--1.05, respectively; cf. Grandi et al. 1997;
Cusumano et al. 1998).
Uncertainties are given at 90\% confidence level
for one interesting parameter, unless otherwise specified.

\subsection{Timing analysis of the persistent emission}

\begin{figure}
\begin{center}
\epsfig{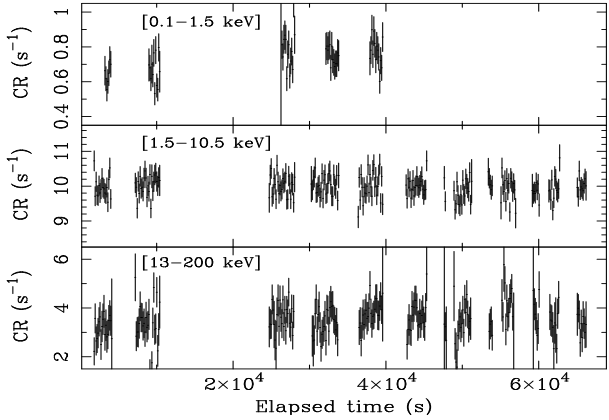}
\end{center}
\caption{Light curves in the 0.1--1.5~keV ({\it upper panel}),
1.5--10.5~keV ({\it central panel}) and 13--200~keV ({\it
lower panel}) energy ranges. The binning time is 128~s.
The LECS and MECS light curves are not background-subtracted.
Only time bins with
an exposure fraction $\ge 25\%$ are shown. Data from the two orbits after a
Type~I burst (corresponding to 12800 to 24200~s) have been removed}
\label{fig5}
\end{figure}

The source underwent a type I burst starting from $\sim$12825~s
from the beginning of the observation,
with an $e$-folding time of $25 \pm 3$~s.
In this paper we define the ``persistent'' emission
excluding the data from the occurrence of the burst up to the
end of the next orbit. The cut corresponds to $\simeq$10$^4$~s
of elapsed time.
Fig.~\ref{fig5} shows 
light curves of the so defined persistent emission
in the 1.5--10.5~keV (MECS), 0.1--1.5~keV (LECS)
%-----------------------------Figure 5--------------------------------
%-----------------------------Figure 5--------------------------------
and 13--200~keV (PDS) energy ranges.
These energy ranges have been chosen in order to sample
different spectral components (see Sect.~3.3).
The $\chi^2$ of a
constant fit for these light curves
is 245 for 220~degrees of freedom (dof), 
96.5/62~dof and 252/210~dof,
respectively. At energies $\simgt 1.5$~keV there is no evidence
of variability, while the relatively high $\chi^2$ obtained from the
light curves at energy $\le$1.5~keV is mainly due to an intensity increase 
of about 15\% between the pre--burst and the post--burst phases.
However, we will in the following analyze the time--averaged spectra
of all instruments. In Sect~3.4 we show that the induced systematic
uncertainties are smaller that the statistical errors
on the spectral parameters, even in the simplest parameterization.

\subsection{The burst}

\begin{figure}
\begin{center}
\epsfig{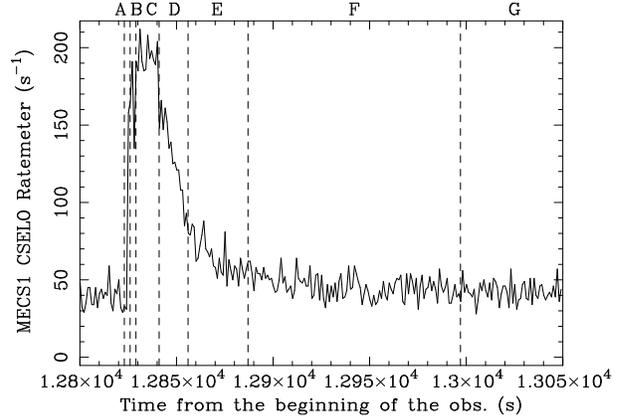}
\end{center}
\caption{Zoom of the MECS {\sc CSELO} ratemeter light curve
around the X1724-308 burst. Dashed vertical lines indicates
the contiguous time intervals used for the time-resolved
burst spectroscopy. For sake of clarity only a fraction of the
``G'' phase is shown, which extends for $\simeq 400$~s
from ${\rm t \simeq 13000}$~s}
\label{figj}
\end{figure}

A profile of the burst in the full MECS bandpass is shown in
Fig.~\ref{figj}. The ratemeter {\sc CSELO} (which registers
%-----------------------------Figure 5--------------------------------
%-----------------------------Figure 5--------------------------------
the total count rate above a low energy threshold) is used to
preserve the continuity of the burst profile in the plot, since
some of the event scientific packets at the peak of the burst were missing
due to telemetry losses. The burst shows a typical exponential
decay with e-folding time ${\rm 25 \pm 3}$~s.
We have performed a spectral analysis of the burst in 7
contiguous time intervals (labeled as
phases ``A'' to ``G'' in the following).
The borders of the intervals are shown as
dashed vertical lines in Fig.~\ref{figj}. The
LECS and HPGSPC were switched off during most of the burst and
PDS detection is generally too faint to allow time-resolved
spectroscopy on these small time scales. Therefore, spectra
of the MECS data only have been extracted, except for
the phases ``F'' and ``G'', when LECS data
were available. The average persistent
spectrum has been used as background.
Since the persistent emission in the 2--10~keV band is
constant on timescales $\sim$10$^2$~s within $\pm5\%$,
time-resolved burst spectra were extracted under
the condition that that 2--10~keV count rate was
$\simgt$20\% of the persistent one. At lower count
rates, the uncertainties on the true instantaneous level
of the persistent emission make the determination
of the spectral parameters meaningless.
The spectra have
been fitted with a simple blackbody emission, photoelectrically
absorbed by a column density, ${\rm N_H}$, of neutral matter.
All the parameters were allowed to vary, except the
absorbing column density in the phase ``C'' spectral fit,
which turned out to be unconstrained and was therefore
fixed to the average of the best--fit values in phases
``E'' and ``F'' and ``G''.
Best--fit parameters and results are reported in Table~\ref{tab6}.
%-----------------------------Table 6--------------------------------
\begin{table*}[hbt]
\caption{Best-fit parameters and $\chi^2$ values when a simple
photoelectric absorbed blackbody model is applied to the time-resolved
spectra of the X1724-308 burst. ${\rm T_{exp}}$ is the exposure time,
${\rm CR}$ the (persistent emission subtracted)
count rate in the 2--10~keV MECS bandpass,
${\rm R_{bb}}$ is the radius of
a spherical blackbody emitting region and ${\rm L_X}$ the total X-ray
luminosity. Errors on CR correspond to 1-$\sigma$.}
\begin{center}
\begin{tabular}{lccccccc} \hline 
Phase & ${\rm T_{exp}}$ & ${\rm CR}$ & N${\rm_H}$ & kT &  ${\rm R_{bb}}$ & ${\rm L_X}$ & ${\rm \chi^2/}$dof \\
& (s) & (s$^{-1}$) & $10^{22}$~cm$^{-2}$ & (keV) & (km) & ($10^{38}$~erg~s$^{-1}$) & \\ \hline
A & 2.8 & $149 \pm 8$ & $< 6.3$ & $3.1\pm^{1.0}_{0.5}$ & $5 \pm^2_4$ & $3.5 \pm^{1.9}_{0.7}$ & 4.6/9 \\
B & 1.17 & $420 \pm 20$ & $< 4.0$ & $2.2\pm^{0.3}_{0.2}$ & $13 \pm^3_4$ & $5.4 \pm^{1.0}_{0.7}$ & 6.6/10 \\
C & 0.79 & $480 \pm 20$ & 1.2$^{\dag}$ & $1.93 \pm^{0.20}_{0.18}$ & $18 \pm 4$ & $6.0 \pm 0.2$ & 5.5/8 \\
D & 3.9 & $250 \pm 8$ & $< 1.7$ & $2.1 \pm^{0.3}_{0.2}$ & $11 \pm 3$ & $3.1 \pm^{0.4}_{0.2}$ & 25/20 \\
E & 31.7 & $78.0 \pm 1.7$ & $1.4 \pm 0.5$ & $1.36 \pm 0.06$ & $12.5 \pm^{1.2}_{1.1}$ & $0.69 \pm 0.04$ & 41.2/40 \\ 
F & 104.7 & $18.8 \pm 0.5$ & $1.3 \pm^{0.9}_{0.7}$ & $1.08 \pm 0.08$ & $9.2 \pm^{1.5}_{1.4}$ & $0.151 \pm^{0.016}_{0.013}$ & 60.3/40 \\
G & 401.9 & $3.22 \pm 0.19$ & $0.8 \pm^{0.8}_{0.5}$ & $0.87 \pm^{0.11}_{0.10}$ & $5.8 \pm^{1.4}_{1.5}$ & $0.025 \pm 0.004$ & 65.9/65 \\ \hline
\end{tabular}
\end{center}
\noindent
$^{\dag}$fixed
\label{tab6}
\end{table*}
%-----------------------------Table 1--------------------------------
The blackbody temperature decreases along the burst profile,
as typically observed in this kind of burst.
The apparent blackbody radius (obtained assuming spherical geometry)
is consistent
with being the same in all phases within the statistical uncertainties,
except for the first (``A'') and the latest (``F'')
phases, for which the best-fit
is $\simeq 5-6$~km. While the value obtained in the former phase
is likely to be
strongly affected by the mixing of blackbodies with a rapidly
changing distribution of temperatures during the sharp rise of the
burst (and therefore probably unreliable), the latter could
suggest a true decrease of the emitting region typical
size with time; this suggestion is supported by the fact
that the nominal best-fit value in phase ``F'' is indeed lower than
between phases ``B'' to ``E''. However, better statistical
quality of the data is required to confirm such a conclusion.
The weighted average of the ${\rm R_{bb}}$
is ${\rm <R_{bb}> = 9.9 \pm 0.7}$~km (${11.6 \pm 0.8}$~km if
only phases between ``B'' and ``F'' are considered).
The peak flux is ${\rm \simeq 5.0 \times 10^{-8}}$~erg~cm$^{-2}$~s$^{-1}$,
corresponding to a luminosity ${\rm \simeq 6 \times 10^{38}}$~erg~s$^{-1}$.
The simple parameterization adopted
yields an adequate fit for all spectra.

\subsection{Spectral analysis of the persistent emission}

We performed the spectral analysis of the persistent emission
in several steps.
First LECS/MECS and HPGSPC/PDS spectra were fitted separately with
purely phenomenological models.
The spectra of all detectors were then fitted
simultaneously, assuming different physical scenarios. 

\begin{figure}
\begin{center}
\epsfig{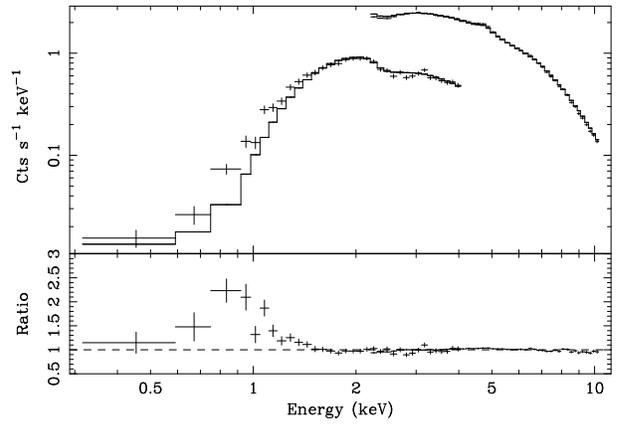}
\end{center}
\caption{LECS and MECS spectra ({\it upper panel}) and data/model
ratio ({\it lower panel}) when 
a photoelectric absorbed power--law is applied}
\label{figa}
\end{figure}

A single component model
with absorption provides
a rather poor description of the LECS/MECS
spectrum. Power--law, bremsstrahlung and blackbody spectral models were
tried. All of these give large excesses above the extrapolations of the 
high-energy data at $\sim$1~keV (see Fig.~\ref{figa}).
%-----------------------------Figure A--------------------------------
%-----------------------------Figure A--------------------------------
A two--component model is needed to account for the observed spectrum
(see Table~\ref{tab1}).
%-----------------------------Table 1--------------------------------
\begin{table*}[hbt]
\caption{Best-fit parameters and $\chi^2$ values when the models in the first
column are applied simultaneously to the LECS and MECS spectra. In column
1 or in the subscripts: {\it wa}~=~photoelectric absorption;
{\it po}~=~power--law; {\it bb}~=~blackbody; {\it brems}~=~thermal
bremsstrahlung.}
\begin{center}
\begin{tabular}{lcccc} \hline 
Model & N${\rm_H}$ & $\Gamma$ or kT${\rm _{brems}}$ & 
T${\rm _{brems}}$ or T${\rm_{bb}}$ or E${\rm_{break}}$ & $\chi^2/$~dof \\
& ($10^{22}$~cm$^{-2}$) & (keV) & (keV) & \\ \hline
{\verb!wa*po!} & $1.86 \pm 0.05$ & $1.834 \pm^{0.015}_{0.0013}$ & \dots 
& 303.4/76 \\
{\verb!wa*bknpo!} & $1.42 \pm^{0.06}_{0.07}$ & $1.55\pm^{0.04}_{0.06}$ & 
$4.3\pm^{0.3}_{0.2}$ & 108.4/74 \\
& & $1.91\pm ^{0.03}_{0.02}$ & \\
{\verb!wa*(po+bb)!} & $1.31 \pm 0.08$ & $1.61\pm 0.05$ 
& $1.08\pm ^{0.05}_{0.04}$ & 89.5/74 \\
{\verb!wa*(po+brems)!} & $2.39\pm 0.09$ & 
$1.924 \pm 0.018$ & $0.18\pm 0.02$ & 101.6/74 \\
{\verb!wa*brems!} & $1.38 \pm 0.04$ & $11.6\pm ^{0.2}_{0.3}
$ & \dots & 123.5/76 \\
{\verb!wa*(brems+bb)!} & $1.46\pm ^{0.06}_{0.05}$ & $11.2\pm^{0.3}_{0.2}$ 
& $0.10\pm 0.03$ & 104.0/74 \\ \hline 
\end{tabular}
\end{center}
\label{tab1}
\end{table*}
%-----------------------------Table 1--------------------------------
Formally, the best combination is a relatively flat (${\rm \Gamma \simeq
1.6}$) power-law + a blackbody with ${\rm kT \simeq 1}$~keV. Other
combinations yield a null hypothesis likelihood ${\rm \simeq 1\%}$.

\begin{figure}
\begin{center}
\vspace{2.5cm}
\epsfig{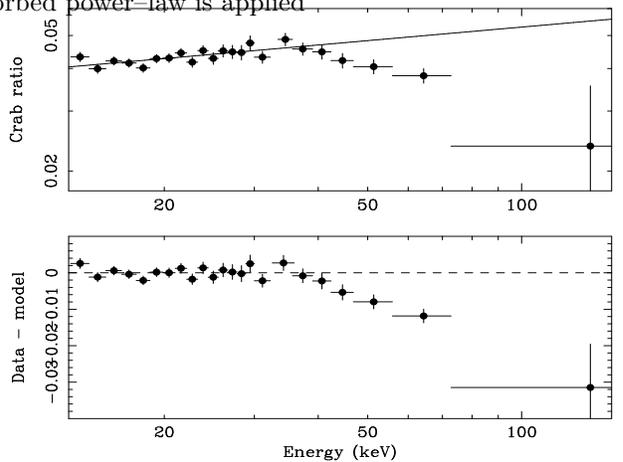}
\end{center}
\caption{Ratio of the PDS spectra of Terzan~ 2 and Crab ({\it upper panel})
and difference residuals ({\it lower panel}) when a power--law model is
applied to fit the ratio in the 13--30~keV band and extrapolated to higher
energies ({\it solid line}).
Each data point has a S/N ratio $\ge$ 20}
\label{figb}
\end{figure}

\begin{figure}
\begin{center}
\epsfig{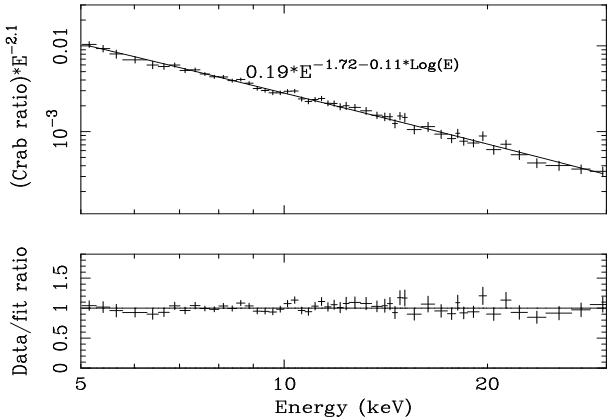}
\end{center}
\caption{In the {\it upper panel} the
ratio of the HPGSPC spectra of X1724-308 and the Crab
Nebula multiplied by $E^{-2.1}$ (typical value of
Crab Nebula photon index, see {\it e.g.}
Cusumano et al., 1998).
The solid line is the best--fit
with the energy dependency law: $0.19 \times E^{-1.72 -0.11 \log(E)}$.
The residuals in ratio between the normalized Crab ratio spectrum
and the best-fit law above are shown in the {\it lower panel}}
\label{figh}
\end{figure}

\src\ is detected in the PDS up to
$\simeq 150$~keV.
One--component models give similarly
poor descriptions of the HPGSPC and PDS spectra (see Table~\ref{tab2}).
%-----------------------------Table 2--------------------------------
\begin{table*}[hbt]
\caption{Best-fit parameters and $\chi^2$ values when the models in the first
column are applied simultaneously to the HPGSPC and PDS spectra. {\it cutoffpl}~=~power--law+
cutoff.}
\begin{center}
\begin{tabular}{lcccc} \hline 
Model & $\Gamma$ & kT${\rm_{br}}$ & E${\rm_{cutoff}}$ & $\chi^2/$~dof \\
& & (keV) & (keV) & \\ \hline
{\verb!po!} & $2.104\pm 0.014$ & \dots & \dots & 266.6/131 \\
{\verb!brems!} & \dots & $28.2 \pm^{0.9}_{0.8}$ & \dots & 667.6/131 \\
{\verb!cutoffpl!} & $1.86\pm 0.04$ & \dots & $99\pm^{19}_{15}$ & 137.6/130 \\
{\verb!po+brems!} & $2.14 \pm 0.08$ & $30 \pm^9_8$ & \dots & 171.7/129 \\ 
\hline 
\end{tabular}
\end{center}
\label{tab2}
\end{table*}
%-----------------------------Table 2--------------------------------
There is strong evidence of a sharp roll over of the PDS spectrum above
$\simeq 40$~keV (see Fig.~\ref{figb}). The inclusion
%-----------------------------Figure B--------------------------------
%-----------------------------Figure B--------------------------------
of a cutoff in the power--law model significantly improves the
quality of the fit ($\Delta \chi^2 = 129$, $\chi^2_{\nu} = 1.06$),
yielding best-fit
parameters of $\Gamma = 1.86 \pm 0.04$ and 
${\rm E_{cutoff}} = 90\pm^{19}_{15}$~keV.
There is evidence in the HPGSPC data
that the spectral index undergoes a smooth
increase with energy (see Fig.~\ref{figh})
%-----------------------------Figure H--------------------------------
%-----------------------------Figure H--------------------------------
from $\Gamma \simeq 1.80$ at 5~keV to $\Gamma \simeq 1.90$ at
30~keV.

\subsubsection{Comparing physical scenarios}

Three spectral region can be phenomenologically recognized
in the \src\ broad-band BeppoSAX spectrum:
an intermediate-hard ({\it i.e.}: 1.5--50~keV) power--law 
``emission core'', with spectral index $\simeq 1.6-1.9$
(probably gently steepening with energy);
a sharp cutoff above $\simeq 50$~keV and a soft excess
at energies below $\simeq 1$~keV.

The models proposed so far to explain the X-ray emission from 
low-mass X-ray binaries (LMXRB)
can be categorized in two main classes. The first one
(originally proposed by Mitsuda et al. 1984) postulates the
superposition of a multicolor blackbody emission from the inner region
of an accretion disk (typical effective temperature $T_{disk} \sim$1~keV) and
of a simple blackbody originating on the surface of the neutron star
($T_{bb} \sim$2~keV).
Another class invokes unsaturated Comptonization of low-energy photons
(White et al. 1985, 1986), possibly coupled
with a blackbody
(White et al. 1988). It has been recently suggested
that these models characterize different luminosity states in LMXRB
(see {\it e.g.} Tanaka 1995), the former prevailing for ${\rm L_X \simgt
10^{37}}$~erg~s$^{-1}$. Actually, the double blackbody
scenario is ruled out by the present data,
at least in its simplest form. 
The fit with such a model yields a very poor
$\chi^2$ (${\rm \chi^2_{\nu}} = 21$); the proposed
model does not produce enough flux at high energy
to account for the relative hardness of
the observed spectrum.

\begin{figure}
\begin{center}
\epsfig{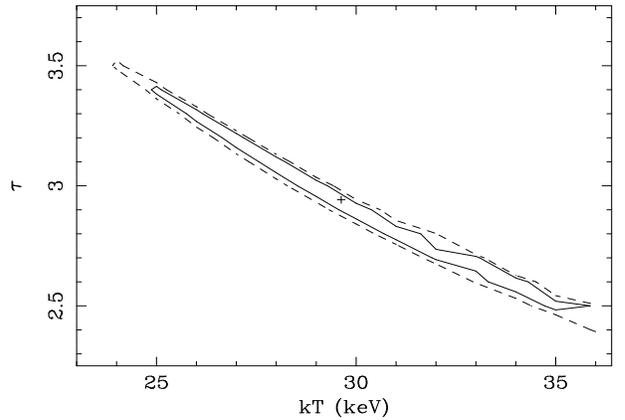}
\end{center}
\caption{Contour plot
$\tau$ vs. kT$_e$ for the model \#5 in Table~\ref{tab3}.
Iso-$\chi^2$ curves are at 90\% and 99\% level of
confidence for two interesting parameters}
\label{figd}
\end{figure}

\begin{figure*}
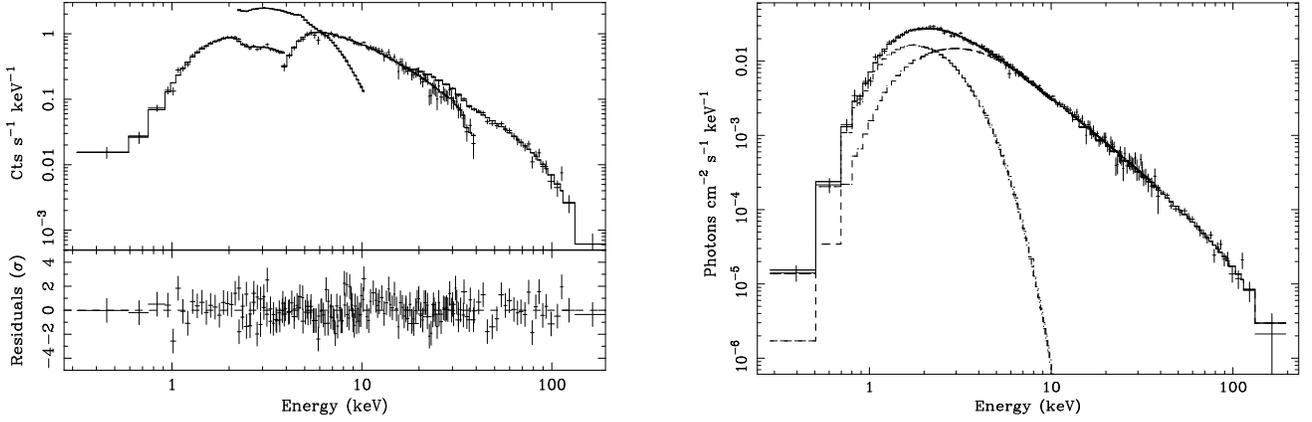

\begin{center}
\epsfig{figure=H1063.f7,height=8.0cm,width=5.5cm,angle=-90}
\hspace{1.cm}
\epsfig{figure=H1063.f8,height=8.0cm,width=5.5cm,angle=-90}
\end{center}
\caption{{\it Left}: LECS, MECS, HPGSPC and
PDS spectra of \src\ ({\it upper panel}) and
residuals in units of standard deviations ({\it lower panel})
when a the best-fit model \#5 in Table~3 is applied.
Each data points has a S/N ratio $>3$. {\it Right}: the inferred
photon spectrum with the best--fit model, deconvolved in the
Comptonized ({\it dashed line}) and disk ({\it dot-dashed line})
blackbody components}
\label{fige}
\end{figure*}

We have tested the viability of the Comptonization scenario
with the {\sc xspec} model {\verb!compTT!} (Titarchuk 1994;
Hua \& Titarchuk 1995; Titarchuk \& Lyubarskij 1995),
which calculates self-consistently the spectrum produced
by the Comptonization of soft photons in a hot plasma. This model
contains as free parameters the temperature of the Comptonized
electrons T${\rm _e}$, the plasma optical depth $\tau$ and the input 
temperature of the soft photon (Wien) distribution T${\rm _W}$. 
Both a disk and a spherical geometry
for the Comptonizing plasma were tried. Again, a soft excess
remains around $\simeq 1$~keV if {\verb!compTT!} alone is applied. We
have tried to model this excess either with a single temperature
or with a multi temperature blackbody, for each of the two plasma geometries.
The soft excess cannot instead be modeled with a Gaussian line, even if
its intrinsic width is allowed to be broad.
The results are summarized in Table~\ref{tab3}.
%-----------------------------Table 3---------------------------------
\begin{table*}
\caption{Best-fit results when Comptonization models are applied
simultaneously to the LECS, MECS and PDS spectra.
{\it diskbb}~=~multi-temperature accretion disk
{\it compTT}~=~Comptonization. For the
{\it compTT} model the letter in bracket indicates the disk (``d'') or
spherical (``s'') geometry for the Comptonizing plasma.
y is the Comptonization parameter (${\rm\equiv 4[kT_e/m_e c^2]} 
\tau^2$]) and R${\rm _W}$ is the typical size of the input photons emitting 
region (following Titarchuk 1994)}

\begin{tabular}{llccccccccc} \hline 
\# & Model & N${\rm _H}$ & kT${\rm _e}$ & $\tau$ & y & kT${\rm _W}$ 
& R${\rm _W}$ & kT${\rm _{se}}$ & R & $\chi^2$/dof \\
& & (10$^{22}$~cm$^{-2}$) & (keV) & & & (keV) & (km) & (keV) & (km) & \\ \hline
1 & {\verb!wa*(compTT!}(d){\verb!)!} & $2.06 \pm^{0.05}_{0.04}$ & $24\pm^{4}_{2}$ & $1.42\pm^{0.16}_{0.17}$ & 0.4 & $0.15 \pm ^{0.02}_{0.03}$ & 350 & \dots & \dots & 518.6/203 \\
2 &{\verb!wa*(compTT!}(d){\verb!+bb)!} & $0.78\pm^{0.10}_{0.07}$ & $29\pm^{18}_{3}$ & $1.17\pm^{0.17}_{0.48}$ & 0.3 &$1.02\pm^{0.13}_{0.12}$ & 7.9 & $0.60\pm^{0.07}_{0.06}$ & $12 \pm^4_3$ & 199.1/201 \\
3 & {\verb!wa*(compTT!}(d){\verb!+diskbb)!} & $1.11\pm^{0.06}_{0.04}$ & $28\pm^{13}_{5}$ & $1.3 \pm^{0.2}_{0.4}$ & 0.4 & $2.10\pm^{0.18}_{0.26}$ & 1.8 & $1.50\pm^{0.09}_{0.13}$ & $2.5\pm^{0.4}_{0.2}$$^{\dag}$ & 194.4/201 \\
4 & {\verb!wa*(compTT!}(s){\verb!)!} & $0.48\pm^{0.04}_{0.03}$ & $23\pm^{4}_{2}$ & $3.7 \pm^{0.3}_{0.4}$ & 2.5 & $0.720\pm^{0.010}_{0.009}$ & 15 & \dots & \dots & 337.6/203 \\
5 & {\verb!wa*(compTT!}(s){\verb!+bb)!} & $0.78\pm^{0.09}_{0.07}$ & $30\pm^{8}_{5}$ & $2.9\pm^{0.5}_{0.6}$ & 2.0 & $1.02 \pm^{0.13}_{0.10}$ & 8.3 & $0.60\pm^{0.06}_{0.08}$ & $12\pm3$ & 198.5/201 \\
6 & {\verb!wa*(compTT!}(s){\verb!+diskbb)!} & $1.13\pm0.04$ & $27\pm^{12}_{4}$ & $3.3\pm^{0.5}_{0.7}$ & 2.3 & $1.98\pm^{0.32}_{0.10}$ & 2.1 & $1.44\pm^{0.16}_{0.08}$ & $2.7\pm^{0.3}_{0.4}$$^{\dag}$ & 194.9/201 \\ \hline
\end{tabular}
\noindent
$^{\dag}$$R = R' \cos(\theta)^{1/2}$, where $\theta$ is the inclination angle and
${\rm R'}$ is the inner radius of the disk
\label{tab3}
\end{table*}
%-----------------------------Table 3---------------------------------
The above scenarios cannot be discriminated on purely statistical
grounds, since the values of $\chi^2_{\nu}$ are all very close to one. 
The Comptonized
plasma is characterized by an electron temperature $\simeq$25--35~keV
(90\% confidence level for two interesting parameters, see
Fig.~\ref{figd})
and a geometry-independent parameters $\beta \sim 0.2$, which
%-----------------------------Figure D--------------------------------
%-----------------------------Figure D--------------------------------
expresses the probability of photon scattering through the exponential
factor $e{\rm^{-\beta}}$ (Titarchuk 1994). The other fit parameters are
slightly dependent on the assumed model used to describe the soft X-ray
excess, typical values being
N${\rm_H} \simeq 0.8$~(1.1)$~ \times 10^{22}$~cm$^{-2}$,
kT${\rm_W} \simeq 1$~(2)~keV, kT${\rm_{se}} \simeq 0.6$~(1.5)~keV 
and R~$\simeq
12$~(3)~km, when it is described with a single- (multi-) temperature
blackbody. The best-fit model \#5 in Table~\ref{tab3} superposed
to the observed spectra is shown in Fig.~\ref{fige}. It corresponds
%-----------------------------Figure E--------------------------------
%-----------------------------Figure E--------------------------------
to the following fluxes:
$6.4 \times 10^{-10}$~erg~cm$^{-2}$~s$^{-1}$ (2--10~keV),
$10.3 \times 10^{-10}$~erg~cm$^{-2}$~s$^{-1}$ (1--20~keV),
$4.7 \times 10^{-10}$~erg~cm$^{-2}$~s$^{-1}$ (38--200~keV),
$18.3 \times 10^{-10}$~erg~cm$^{-2}$~s$^{-1}$ (0.1--200~keV),
the last one corresponding to an {\it unabsorbed} X-ray luminosity
L$_X \sim 2.3 \times 10^{37}$~erg~s$^{-1}$ for a distance of 10~kpc
(Djorgovski 1993). Given such a continuum, the 90\% upper limit
on the equivalent width of a narrow line
at 6.7~keV is 17~eV.

However, it must be noted that the broadband spectrum can be formally
accounted by a suitable combination of three thermal models, as shown
in Table~\ref{tab4}. A bremsstrahlung with temperature
T${\rm _{br}} \simeq 45-60$~keV is required to fit the PDS hard tail,
a 0.8--1~keV blackbody can account for the soft excess,
while a $1.9$~keV
%-----------------------------Table 4-------------------------------------
\begin{table*}
\caption{Best-fit results for the ``thermal scenarios'' of the 
Terzan 2 broadband spectrum. The subscripts of the kT refer to the thermal 
components in the model in column 1 read from the left to the right. 
R$_{bb}$ is the size of a spherical optically thick blackbody emitting region}
\begin{center}
\begin{tabular}{llcccccc} \hline 
\# & Model & N${\rm_H}$ & kT$_1$ & R${\rm_{bb}}$ & kT$_2$ & kT$_3$ 
& $\chi^2$/dof \\  
& & ($10^{22}$~cm$^{-2}$) & (keV) & (km) & (keV) & (keV) & \\
\hline
7 & {\verb!wa*(bb+bb+brems)!} & $0.95 \pm^{0.05}_{0.04}$ & $0.83 \pm^{0.03}_{0.04}$ & $6.5 \pm^{0.7}_{0.5}$ & $1.98 \pm^{0.16}_{0.13}$ & $48 \pm 3$ & 212.3/201 \\
& & & & $1.17 \pm^{0.16}_{0.19}$ & & & \\
8 & {\verb!wa*(bb+brems+brems)!} & $1.25 \pm ^{0.07}_{0.05}$ & $0.99\pm^{0.05}_{0.04}$ & $2.8 \pm^{0.4}_{0.5}$ & $9.3\pm^{1.5}_{1.3}$ & $62\pm^{9}_{7}$ & 202.2/201 \\ \hline 
\end{tabular}
\end{center}
\label{tab4}
\end{table*}
%-----------------------------Table 4-------------------------------------
blackbody or a $\simeq 9$~keV bremsstrahlung are needed to fill the gap
between the two ``extreme'' spectral components. In both cases, the
$\chi^2$ is as good as for the Comptonization models.

\subsubsection{The soft X-ray variability}

After the burst, the 0.1--1.5~keV flux increased by ${\rm \simeq 15\%}$.
LECS and MECS spectra corresponding
to these intervals were extracted and analyzed with
the same model (model \#5 in Table~\ref{tab3}
with kT${\rm _e} = 28$~keV and $\tau = 3.1$),
in order to investigate if there is any significant spectral variability
associated with this flux change. The best--fit parameters are summarized in
Table~\ref{tab5}. The $\chi^2$ for the combined fit is acceptable
%-----------------------------Table 4-------------------------------------
\begin{table*}
\caption{Best--fit parameters for the joint fit of the LECS and MECS
spectra extracted before and after the burst}
\begin{center}
\begin{tabular}{lcccc} \hline 
Temporal phase & N${\rm_H}$ & kT${\rm_W}$ & kT${\rm_{bb}}$ 
& ${\rm N_{bb}}$ \\
& (10$^{22}$~cm$^{-2}$) & (keV) & (keV) & (${\rm 10^{-3} L_{39 erg s^{-1}} D^{-2}_{10 kpc}}$) \\ \hline
Before the burst & $087\pm^{0.17}_{0.13}$ & $1.1\pm 0.2$ 
& $0.64 \pm^{0.10}_{0.14}$ & $2.6\pm^{0.9}_{1.1}$ \\
After the burst & $0.84 \pm^{0.12}_{0.09}$ & $0.99\pm^{0.13}_{0.11}$ 
& $0.57\pm^{0.07}_{0.06}$ & $2.3 \pm 0.5$ \\ \hline
\end{tabular}
\end{center}
\label{tab5}
\end{table*}
%-----------------------------Table 4-------------------------------------
($\chi^2_{\nu} = 1.13$). All the best-fit parameters are mutually
consistent between the two intervals. We conclude that
the systematic uncertainties introduced by mixing models
with different normalizations are well below the statistical
uncertainties, given the quality of the current data.

\section{Discussion}

We report on the first X-ray broadband spectrum of the burster
X1724-308 in the whole 0.1--100~keV domain. The
source is detected with a S/N $>3$ up to $\sim$150~keV, confirming
the previous Sigma/GRANAT results (Barret et al. 1991). The superior
combination of broadband coverage, sensitivity and energy resolution
has however allowed an unprecedentedly detailed spectral deconvolution,
albeit formally not unique. Qualitatively the spectrum can be
described as the composition of a ``core'', power--law emission,
which undergoes
a significant steepening at E$\simeq 50$~keV, and a soft
excess above the extrapolation of the higher energy X-ray spectrum.
The 20--200~keV flux is
${\rm \sim 3.4 \times 10^{-10}}$~erg~cm$^{-2}$~s$^{-1}$.
The corresponding luminosity at 10~kpc is
${\rm \sim 4.1 \times 10^{36}}$~erg~s$^{-1}$, which is
typical of X-ray bursters (van Paradijs \& van der Klis 1994;
Barret et al. 1996).

Formally, the spectrum can be described by a
suitable combination of three thermal models (models \#7 and \#8 in
Table~\ref{tab4}). The characterizing feature of these models is
the need for a very high temperature thermal bremsstrahlung
(T${\rm_{br}} \sim 50-60$~keV) to account for the highest energy PDS data 
points and
cut-off roll over. The presence of such a component had already
been invoked to explain the hard tail observed in several X-ray bursters
(Churazov et al. 1995). The luminosity
inferred by
the BeppoSAX observation is L${\rm_{br}} \sim 1.4 \times 10^{37}$~erg~s$^{-1}$,
which implies an emission measure
${\rm < n_e V^2 >} \sim 3 \times 10^{59}$~cm$^3$; the optically thin
requirement leads to a size of a single uniform emitting cloud
$>$$7 \times 10^{10}$~cm. Such a large region is unlikely
to be directly powered by the neutron star.

The high-energy tail rules out the ``double-blackbody'' model
(multi-temperature
disk + neutron star surface), which had originally proposed by
Mitsuda et al. (1984) to explain the X-ray emission of LMXRB. If a
transition from a thermal- to a Comptonization-dominated hard X-ray
spectrum is present in \src\ (as generally believed for LMXRB), it
occurs at luminosity ${\rm \simgt}$ a few ${\rm 10^{37}}$~erg~s$^{-1}$.

Models of the BeppoSAX spectrum of \src\ which involve a Comptonized 
component appear therefore more plausible. 
A plasma with kT${\rm _e} \sim$27--30~keV and
$\beta \simeq 0.2$ is
required. A disk and a spherical geometry of the Comptonizing region fit
the data comparably well. We note however, that somewhat higher 
values of the Comptonization parameter ($y \simeq$2.0--2.3) are 
obtained for a spherical geometry.
Similar Comptonized spectral components have been observed in a variety 
of neutron star LMXRBs, although in most cases the spectra are 
substantially softer and the inferred electron temperatures in the 
2--4~keV range (White et al. 1988).
In a few Atoll sources, however, X-ray spectra extending 
up to $\sim$100~keV or more have been revealed through {\it Sigma}/GRANAT 
(Mandrou et al. 1994) and BeppoSAX (Church et al. 1998b)
observations. These spectra have a clear resemblance to the 
spectrum of \src\ discussed here, and require comparably high 
electron temperatures. It has also to be considered that hard 
tails extending to hundreds of keV are no longer believed to be exclusive
to black hole candidates, at least when the LMXRB luminosity
does not exceeds ${\rm \sim}$${ 10^{37}}$~erg~s$^{-1}$ (Barret et al. 1994;
van Paradijs \& van der Klis 1994). 

In the Beppo-SAX observation of \src\, an additional soft spectral
component is needed to fit the low energy spectrum. However, the quality of
the data does not allow different models for this
component to be discriminated.
We find that a multi-temperature blackbody disk model, or a simple
blackbody spectrum provide acceptable fits. 

In the case of a multi-temperature blackbody disk model, 
both the derived temperature ($\sim$2~keV)
and radius of the innermost disk region ($\sim$$3/(\cos \theta)^{1/2}$~km)
are close to those obtained for the ``seed" Wien spectrum for the 
Comptonized component (see {\it e.g.} the model \#6 in Table~\ref{tab4}). 
An inclination angle of $\sim 70^{\circ}$ would be required for the 
inner radius of the disk to accommodate a neutron star. 
On the contrary the absence of a sizeable orbital modulation of the 
X-ray flux (absorption dips in particular) suggests a substantially lower 
inclination (P89; Olive et al. 1998). 

If the soft thermal emission of Terzan 2 is interpreted in terms of a 
single temperature blackbody, the inferred linear size of the 
emitting region ($\simeq 12$~km) is in
good agreement with a typical neutron star and with the size inferred
from the analysis of the type I burst occurred during the BeppoSAX
observation. 
It is worth emphasizing the (single) blackbody temperature of  
Terzan 2 (kT$\sim 0.6$~keV) is a factor of 2-3 lower than that derived from 
most other neutron LMXRBs (cf. White et al. 1988). 
In the single temperature blackbody scenario, the parameters 
also are close to those derived for the input spectrum 
to the Comptonized component, such that the neutron stellar surface might 
represent the source of ``seed" photons. Alternatively, the analogy with
the hard spectra of black hole candidates (where, of course, there is 
no star surface or boundary layer) suggest that even in disks 
around accreting neutron stars there is 
a hot Comptonizing plasma that coexists with a dense cold phase 
responsible for the generation of the ``seed" photons 
(cf. Eardley \& Lightman 1975; Galeev et al. 1979; 
Molendi \& Maraschi 1990).

The two spectral decompositions discussed above differ a great deal 
in the physical interpretation of the two spectral
components. In the former case the Comptonized hard component 
originates from the boundary layer close to the neutron star surface,
whereas the soft component is produced by the sum of 
blackbody spectra emitted by the accretion disk. In the latter 
case the (single) blackbody spectrum likely arises from
the boundary layer close to the neutron star surface, while 
the accretion disk gives rise to the Comptonized spectrum. 
Simple accretion disk theory predicts that, unless the neutron star 
is rotating very close to its break up velocity, the
ratio of the accretion luminosity released in the disk to that released in
the boundary layer should be equal to 1. 
Sunyaev \& Shakura (1986) show
that this ratio can be as low as $\sim 0.45$ if the accretion disk 
ends at the marginally stable orbit and this, in turn, is larger then 
the neutron star radius. Recent observations of fast quasi periodic 
oscillations, QPOs, in LMXRBs testify that the neutron stars contained in
them  rotate much slower than break up (periods in the $\sim$3~ms range),
and that the inner radius of the accretion disk (as inferred from the 
frequency of the higher kHz QPO power spectrum peak) is often larger 
than the marginally stable orbit. 
This would further decrease the predicted ratio of disk and boundary
layer luminosity. The 0.1--100~keV
luminosities of the disk (multi-) blackbody component and the (boundary
layer) Comptonized component inferred from the BeppoSAX spectrum of
Terzan 2 are $\sim$$8.5 \times 10^{36}$~erg~s$^{-1}$ and
$\sim 1.5 \times 10^{37}$~erg~s$^{-1}$.
The derived ratio of $\sim$0.55 is in agreement with the predictions 
above. However, observations of dips in several X-ray
can be convincingly explained if the Comptonized component is
extended (Church \& Baluci\'nska-Church 1995; Church et al. 1998a),
almost at odds with the former interpretation.

In the latter interpretation the ratio of the luminosity 
of the Comptonized (disk) spectrum ($1.9 \times 10^{37}$~erg~s$^{-1}$)
and the luminosity of the (boundary layer) single temperature blackbody 
($2.4\times 10^{36}$~erg~s$^{-1}$) is $\sim 8$, {\it i.e.} much 
higher then expected.
An anomalously high ratio of the two components has also been found in a
number of other neutron star LMXRBs (see White et al. 1988).
Obscuration by an inner accretion corona or a thickened disk
could decrease the relative strength of the boundary layer contribution
(Lamb 1986; van der Klis et al. 1987). In the standard
Shakura-Sunyaev disk, the inner radiation pressure-dominated disk can be
as thick as $\sim$ few tens of kilometers, and the effect of thermal
and/or secular instabilities can further enlarge it (Shakura \&
Sunyaev 1973; Pringle 1981). Radiation from the boundary layer
might be partly scattered out of the polar channel or attenuated
through a not completely optically-thick disk.
A ``bloated-disk'' geometry has already been proposed
to explain the low frequency QPOs observed in several LMXRBs
({\it e.g.}: 4U1820-30, Stella et al. 1987),
and would be in line with the likely low inclination
of the system, as inferred from
the lack of significant intensity eclipses.
Alternatively, exchange of energy between the magnetosphere
and the accretion disk via material torque from the accretion matter could
reduce the luminosity available at the surface of a neutron star, spinning
at a rate close to the inner disk Keplerian frequency (Priedhorsky 1986).

The optical reddening is $A_V = 4.03$ (Djorgovski 1993). If we
convert the visual absorption in an X-ray absorbing column density
with N${\rm _H = 1.8 \times 10^{21}}$~A${\rm _V}$~cm$^{-2}$ (Predehl \& Schmitt
1995), the corresponding column density is roughly
consistent with the Galactic one along the line of sight to Terzan~2
(N${\rm _{H_{Gal}} = 6.5 \times 10^{21}}$~cm$^{-2}$, Dickey \& Lockman 1990).
However, the BeppoSAX observation requires a 20-75\% excess absorption
(depending on the spectral descriptions), in agreement
with previous estimates, although rather poorly constrained and
model dependent (P89; Mereghetti et al. 1995; Verbunt et al. 1995).

EXOSAT observed \src\ in a factor $\sim$5 fainter 
and in a much softer state than BeppoSAX. This is shown in
Fig.~\ref{figc}, where the photon spectra observed by BeppoSAX
%-----------------------------Figure C--------------------------------
\begin{figure}
\begin{center}
\epsfig{figure=H1063.f9,height=7.0cm,width=7.0cm,angle=0}
\end{center}
\caption{Beppo-SAX and EXOSAT count spectra. EXOSAT data
point have been multiplied by 2.5 to align the spectra
below the photoelectric absorption cut-off}
\label{figc}
\end{figure}
%-----------------------------Figure C--------------------------------
and EXOSAT are compared. The hardness difference can be parameterized
as a $\Delta \Gamma \simeq 0.7$ between the observed power law photon
indices. If the electron temperature remains the same, $\beta \sim 0.3$ and
therefore a $\sim$50\% smaller $\tau$ (Hua \&
Titarchuk 1995)
at the EXOSAT epoch is implied
(we note however that
in this regime the analytical approximations
on which the Comptonization Monte-Carlo simulation by Titarchuk
(1994) are no longer valid and therefore the condition that T${\rm_e}$ and
$\beta$ determines univocally the spectral index is not fulfilled,
Hua \& Titarchuk 1995).
Alternatively, a steeper spectrum could imply a lower kT${\rm _e}$
and therefore a spectral cut-off at lower energy. To check
this possibility, we
extracted from the public archive
and re-analyzed the EXOSAT ME spectrum of P89. A simple
absorbed power--law model gives a reasonable fit
(${\rm \chi^2_{\nu}} = 77/61$~dof), with best-fit parameters as in P89.
However, the inclusion of a cut-off results in a
reduction in $\chi^2$ of 8 which
is significant at 98.9\% confidence. The best-fit parameters
are then N$_H = (1.5 \pm 0.2) \times 10^{22}$~cm$^{-2}$ 
(very similar to the BeppoSAX value in model \#6 here), 
$\Gamma = 2.0 \pm 0.2$ and
E${\rm _{cut-off} = 14^{+21}_{-6}}$~keV. We caution that this
cut-off energy is very close to the upper energy range of the ME
detector. If the EXOSAT result is correct, then an explanation 
of the difference in the two spectra may be
a factor $\simeq 4$ lower electron temperature during the EXOSAT
epoch.
In such a scenario the energy of the electrons is physically
linked with the high-energy intensity.

\begin{acknowledgements}
 
The authors acknowledge useful suggestions from D. Barret.
The referee's comments have allowed us to enlarge and deepen the
analytical and interpretative approach of this paper.
The BeppoSAX satellite is a joint Italian--Dutch program.
MG and TO acknowledge the receipt of an ESA Research Fellowship.
LS acknowledges the receipt of an ASI grant.

\end{acknowledgements}

\end{document}